\documentclass[12pt]{iopart}

\usepackage[utf8]{inputenc}
\usepackage{a4wide}

\expandafter\let\csname equation*\endcsname\relax

\expandafter\let\csname endequation*\endcsname\relax

\usepackage{amsmath}
\usepackage{amssymb,amsthm,mathrsfs}
\usepackage{dsfont}
\usepackage{graphicx}

\newcommand{\ket}[1]{|#1\rangle}
\newcommand{\braket}[2]{\langle#1|#2\rangle}
\newcommand{\ketbra}[2]{|#1\rangle\langle #2|}
\newcommand{\braketA}[3]{\langle#1|#2|#3\rangle}

\newcommand{\sgn}{\operatorname{sgn}}

\begin{document}

\title{Monitored Recurrence of a One-parameter Family of Three-state Quantum Walks}

\author{Martin \v Stefa\v n\'ak}

\address{Department of Physics, Faculty of Nuclear Sciences and Physical Engineering, Czech Technical University in Prague, B\v rehov\'a 7, 115 19 Praha 1 - Star\'e M\v esto, Czech Republic}
\ead{martin.stefanak@fjfi.cvut.cz}
\vspace{10pt}
\date{\today}

\begin{abstract}
Monitored recurrence of a one-parameter set of three-state quantum walks on a line is investigated. The calculations are considerably simplified by choosing a suitable basis of the coin space. We show that the Polya number (i.e. the site recurrence probability) depends on the coin parameter and the probability that the walker is initially in a particular coin state for which the walk returns to the origin with certainty. Finally, we present a brief investigation of the exact quantum state recurrence. 
\end{abstract}

\maketitle

\section{Introduction}

Quantum walks \cite{Aharonov1993,Meyer1996,Farhi1998} are generalizations of classical random walks to evolution of a quantum particle on a graph or a lattice. Similar ideas appeared already in 1960's in the works of Feynman and Hibbs on discretization of the Dirac equation \cite{feynman_1965} and in the 1980's in the work of Gudder on quantum graphic dynamics \cite{gudder_quantum_1988,gudder_book_1988}. Quantum walks became versatile tools in various field of physics and computation, for a recent review see \cite{qw_review_2021}.

One of the interesting questions in stochastic processes is the problem of recurrence. For classical random walks, this was studied in detail by Polya \cite{Polya1921} who proved that in 1 and 2 spatial dimensions the balanced walk returns with certainty, i.e. is recurrent, while for $d\geq 3$ the recurrence probability (also called Polya number) is strictly less than unity, implying transience. Nevertheless, walks on infinite line and square lattice are null-recurrent, since the expected recurrence time is infinite. On the other hand, irreducible walks on finite graphs are positive-recurrent since the expected recurrence time is finite. In fact, for finite state space Kac shown \cite{Kac1947} that it is an integer.

Studying recurrence of quantum walks requires a detailed description of the measurement process since it has a nontrivial effect on the wavefunction of the quantum walker. To minimize the influence of measurement we have proposed a scheme \cite{Stefanak2008a,Stefanak2008b} where the quantum walk evolves freely for a certain number of steps, then a measurement detecting the presence of the walker at the origin is performed, after which the process is restarted. In this concept the recurrence or transience of a quantum walk can be decided from the behaviour of the probability of finding the walker at the origin after $t$ steps (without prior measurement) in the same way as for the classical random walk. In fact, for classical random walks this scheme is equivalent to the Polya approach, where we monitor the origin after each step, in the sense that the walk is recurrent in one approach if and only if it is recurrent in the other scheme. This stems from the fact that in the classical setting, the fundamental role is played by probabilities and there is a direct relation between the return probabilities without prior measurement and the first return probabilities of the monitored walk. However, this does not hold anymore in the quantum case. Due to the effect of measurement on the wavefunction the two probabilities cannot be accessed in the same experiment and they are not related to each other. Nevertheless, there exists a relation between the return amplitudes of the unitary quantum walk and the first return amplitudes of the monitored walk, as shown in \cite{grunbaum_recurrence_2013,gruenbaum2014} for a much broader setting of iterated unitary evolutions. Note that some particular examples were investigated before the general theory was developed, e.g. the monitored recurrence of the Hadamard walk on a line \cite{Kiss2009} which was shown to be equivalent to the absorption of the walk on the half-line \cite{Ambainis2001}. In the monitored recurrence approach, the expected return time to the exact initial state (state recurrence) of a finite system is an integer \cite{grunbaum_recurrence_2013} as in the classical case. This holds even for iterated open quantum evolutions \cite{Sinkovicz1,Sinkovicz2}. Note that in the coined quantum walks return to the initial position can be understood as a subspace recurrence, since we are not interested in the exact state of the quantum coin. For finite systems the expected return time for a subspace recurrence is a rational number \cite{gruenbaum2014}. We point out that in the recent years the study of recurrence was considerably broadened to open quantum systems, quantum Markov chains and open quantum walks, see eg. \cite{lardizabal_class_2015,carvalho_site_2016,grunbaum2018,bardet_recurrence_2019,dhahri_open_2019,grunbaum_quantum_2020,jacq_homogeneous_2021}.

In contrast to the classical case the two schemes for detecting recurrence (restart after measurement vs monitored after each step) are not equivalent for quantum walks. As an example, the Hadamard walk on a line is recurrent in the restart scheme \cite{Stefanak2008a} while transient in the monitored case \cite{Kiss2009}. This was also demonstrated in an optical experiment \cite{Nitsche2018} where the quantum walk was simulated by a weak laser pulse cycling in a time-delay loop, utilizing time-multiplexing to encode the position and number of steps of the walk into the time of detection of a photon. It was a considerable experimental challenge to perform the local measurement at the origin without disturbing the rest of the pulse sequence. The deterministic out-coupling from the loop was achieved by a programmable fast-switching electro-optical modulator, which allowed to address specific time slots corresponding to the walker being at the origin. 

In the present paper we take the monitored approach to the recurrence problem and apply it to the three-state quantum walks on a line with a particular one-parameter family of coins \cite{stefanak2012}. We show that there is a unique initial coin state for which the walk stays at the origin in the first step, so in this case the walk is recurrent. We then choose this initial state as one of the basis vectors in the coin space and complement the basis with two vectors from orthogonal complement. This leads to significant reduction of the complexity of the follow-up analytical calculations. We show that the subspace recurrence probability is determined by the coin parameter and the probability that the walker is initially in the unique recurrent state. For the exact quantum state recurrence we provide a numerical investigation. In particular, we identify initial states for which the state recurrence is greater than the subspace recurrence, a paradoxical feature of monitored quantum evolution which was already discussed previously in the literature  \cite{gruenbaum2014}.

The paper is organized as follows. Section~\ref{sec2} reviews the basic concepts and tools for investigation of monitored recurrence. Section~\ref{sec3} is dedicated to the study of site recurrence for a one-parameter family of three-state quantum walks. In Section~\ref{sec4} we present a numerical investigation of the state recurrence. We conclude and present an outlook in Section~\ref{sec5}. 

\section{Monitored site recurrence of a quantum walk}
\label{sec2}

We begin by a brief overview of the methods for studying recurrence developed in \cite{grunbaum_recurrence_2013,gruenbaum2014}, adopted for a coined quantum walks with constant coin. Consider a discrete time quantum walk on a lattice starting from the origin. The Hilbert space is a tensor product of the position and the coin spaces
$$
\mathcal{H} = \mathcal{H}_p\otimes\mathcal{H}_c.
$$
We denote the evolution operator by $U$, it has the usual decomposition 
\begin{equation}
\label{evol:op}
U = S\cdot (I_p\otimes C),    
\end{equation}
where $S$ is the shift operator and $C$ is the coin. Return of the quantum walker to the original site can be understood as a subspace recurrence \cite{gruenbaum2014}, since we are not interested in the internal state of the coin, only the position of the quantum walker. Let us denote the orthogonal projector onto the origin subspace as 
\begin{equation}
\label{pi0}
 \Pi_0 = \ketbra{0}{0}\otimes I_c.   
\end{equation}
The evolution of the monitored quantum walk is given by the operator
$$
\tilde{U} = (I-\Pi_0) U,
$$
i.e. the walk continues if we do not find the walker at the origin. Normalized state after $n$ steps of the monitored walk reads
$$
\ket{\psi_n} = \frac{1}{\sqrt{s_n}} \tilde{U}^n\ket{\psi}, \quad \ket{\psi} = \ket{0}\otimes\ket{\psi_c},
$$
where $s_n$ is the survival probability until the $n$-th step
$$
s_n = \lVert \tilde{U}^n\psi\rVert^2.
$$
$s_n$ is the probability that the walker has not returned to the origin in the first $n$ steps. Hence, the complement of the limiting value of $s_n$
$$
P(\psi) = 1 - \lim\limits_{n\to\infty} s_n,
$$
corresponds to the probability that the walker ever returns, i.e. the site recurrence probability (or Polya number). 
Alternatively, we can derive the Polya number from the first return probabilities. Let us denote the probability of first return to the origin after $n$ steps as $q_n$. Since these events are mutually exclusive, the overall site recurrence probability is given by the sum
$$
P(\psi) = \sum_{n=1}^\infty q_n .
$$
In case of iterated unitary evolution the first return probability after $n$ steps is given by
$$
q_n = \lVert \Pi_0 U \tilde{U}^{n-1}\psi\rVert^2 = \lVert a_n\psi\rVert^2 ,
$$
where we have introduced the first return amplitude operator (note that $\Pi_0 \psi = \psi$)
$$
a_n = \Pi_0 U \tilde{U}^{n-1}\Pi_0 . 
$$
The Polya number of a quantum walk for a given initial state is therefore given by
$$
P(\psi) = \sum_{n=1}^\infty \lVert a_n\psi\rVert^2 .
$$

Determining the operators $a_n$ directly is rather difficult, however, they can be related to the $n$-th step return amplitude operators without prior monitoring, which we denote as
$$
\mu_n = \Pi_0 U^n \Pi_0 . 
$$
We define the operator valued generating functions (for complex variable $z$ with $|z|< 1$)
$$
\mu(z) = \sum_{n=0}^\infty \mu_n z^n , \quad a(z) = \sum_{n=1}^\infty  a_n z^n . 
$$
Note that $\mu(z)$ is also called the Stieltjes operator and the first return generating function $a(z)$ is related to the Schur function $f(z)$ by
$$
a(z) = z f^\dagger(z),
$$ 
which is more extensively used in the literature \cite{grunbaum_recurrence_2013,gruenbaum2014,grunbaum2018,grunbaum_quantum_2020}. 
Introducing the resolvents
$$
G(z) = \sum_{n=0}^\infty U^n z^n = (I - z U)^{-1},\quad \tilde{G}(z) = \sum_{n=0}^\infty \tilde{U}^n z^n = (I - z \tilde{U})^{-1} ,
$$
we see that the generating functions can be written in the form  
\begin{equation}
\label{ampl:res}
\mu(z) = \Pi_0 G(z) \Pi_0, \quad a(z) = z \Pi_0 U \tilde{G}(z) \Pi_0 . 
\end{equation}
Using the resolvent identities
$$
G(z) - \tilde{G}(z) = z G(z) \Pi_0 U \tilde{G}(z) = z \tilde{G}(z) \Pi_0 U G(z), 
$$
we can derive the renewal equations \cite{grunbaum_recurrence_2013,gruenbaum2014}
\begin{equation}
\label{renew:eq}
\mu(z) a(z) = a(z) \mu(z) = \mu(z) - \Pi_0 .   
\end{equation}
In the above equation all operators act on the origin subspace (see (\ref{pi0}) and (\ref{ampl:res})), i.e. they are of the form
$$
\mu(z) = \ketbra{0}{0}\otimes \mu_c(z), \quad a(z) = \ketbra{0}{0}\otimes a_c(z),
$$
where $\mu_c(z)$ and  $a_c(z)$ act on the coin space.  Thus, we can rewrite (\ref{renew:eq}) into operator equation on the coin space
\begin{equation}
\label{ac}
a_c(z) = I_c - \mu_c(z)^{-1}.
\end{equation}
Hence, we can express the first return generating function $a_c(z)$ from the return generating function $\mu_c(z)$, which is easier to obtain. All functions can be extended to the unit circle in the complex plane by the radial limit. The site recurrence probability is expressed in terms of the boundary values \cite{grunbaum_recurrence_2013,gruenbaum2014} by 
\begin{equation}
P(\psi_c) = \int\limits_0^{2\pi} \lVert a_c(e^{i t})\psi_c\rVert^2 \frac{dt}{2\pi} = \langle \psi_c|R|\psi_c\rangle,
\label{polya}
\end{equation}
where we have denoted the recurrence probability operator
\begin{equation}
\label{rec:op}
R = \int\limits_0^{2\pi} a_c^\dagger (e^{i t}) a_c(e^{i t}) \frac{dt}{2\pi} .    
\end{equation}
In summary, the recipe to obtain the Polya number is to find the Stieltjes operator $\mu_c(z)$, from (\ref{ac}) we find the first return generating operator, construct the recurrence probability operator (\ref{rec:op}) and find its average value for a given initial state of the coin $\psi_c$ (\ref{polya}). 

The site recurrence probability of a two-state quantum walk on a line, where the walker can move to the right or left in each step, was evaluated explicitly in \cite{grunbaum_recurrence_2013}. The unitary coin operator was parameterized in the following way
\begin{equation}
\label{coin:2state}
C = \begin{pmatrix}
\rho & -\gamma \\
\overline{\gamma} & \rho
\end{pmatrix}, \quad \rho = \sqrt{1-|\gamma|^2} .     
\end{equation}
Since the considered walk is translationally invariant, one can  use Fourier transformation to diagonalize the step operator and evaluate the Stieltjes operator with the formula
\begin{equation}
\label{muc:int}
    \mu_c(z) = \int\limits_0^{2\pi} \frac{dp}{2\pi} (I_c - z U(p))^{-1}.
\end{equation}
Here $U(p)$ is the evolution operator (\ref{evol:op}) in the Fourier picture
$$
U(p) = S(p)\cdot C = \begin{pmatrix}
e^{i p} & 0 \\
0 & e^{-i p}
\end{pmatrix} \cdot \begin{pmatrix}
\rho & -\gamma \\
\overline{\gamma} & \rho
\end{pmatrix},
$$
which is a multiplication operator. Using the substitution $e^{ip} = x $, $ dp = \frac{dx}{ix}$ the RHS of (\ref{muc:int}) is turned into an integral over the unit circle in the complex plane, which can be evaluated with the residue theorem. After some calculations it is found that the recurrence operator (\ref{rec:op}) for a two-state walk is a multiple of identity, i.e. the Polya number does not depend on the initial coin state, only on the quantum coin. 
Explicitly, the site recurrence probability for a two-state quantum walk on a line with the coin (\ref{coin:2state}) is given by \cite{grunbaum_recurrence_2013,sabri_2018}
$$
P = \frac{2\left(|\gamma|\sqrt{1-|\gamma|^2} - (1-2|\gamma|^2)\arcsin|\gamma|\right)}{\pi|\gamma|^2} .
$$

\section{Site recurrence of a three-state walk on a line}
\label{sec3}

Let us now consider a three-state walk on a line, where the walker can move to the right, stay, or move to the left. This corresponds to the standard basis of the coin space $\ket{R}$, $\ket{S}$ and $\ket{L}$. We choose the following one-parameter set of coins \cite{stefanak2012,stefanak2014}
\begin{equation}
\label{coin}
C = \begin{pmatrix}
-\rho^2 & \rho\sqrt{2(1-\rho^2)} & 1-\rho^2 \\
\rho\sqrt{2(1-\rho^2)} & 2\rho^2-1 & \rho\sqrt{2(1-\rho^2)} \\
1-\rho^2 & \rho\sqrt{2(1-\rho^2)} & -\rho^2
\end{pmatrix} ,  \quad 0<\rho<1,  
\end{equation}
which reduces to the 3x3 Grover matrix for $\rho=\frac{1}{\sqrt{3}}$. Before turning to the derivation of the Polya number, we make the following observation. Since the coin operator (\ref{coin}) is equal to its inverse, the initial coin state 
$$
\ket{\alpha_1} = C\ket{S} = \rho\sqrt{2(1-\rho^2)} (\ket{R} + \ket{L}) + (2\rho^2-1)\ket{S} =  \begin{pmatrix}
\rho\sqrt{2(1-\rho^2)}\\
2\rho^2-1 \\
\rho\sqrt{2(1-\rho^2)}
\end{pmatrix},
$$
is mapped to $\ket{S}$. Hence, if we take $\ket{\alpha_1}$ as the initial coin state, the walker is absorbed with certainty after one step, i.e. for this initial state the site recurrence probability is one. Numerical simulations indicate that for all coin states in the orthogonal complement to $\ket{\alpha_1}$ the Polya number has the same value, which is less than unity. Let us complete the orthonormal basis in the coin space by
\begin{align*}
\ket{\alpha_2} & =  \frac{2\rho^2-1}{\sqrt{2}}(\ket{R} + \ket{L}) - 2\rho\sqrt{1-\rho^2} \ket{S} = \begin{pmatrix}
\frac{2\rho^2-1}{\sqrt{2}} \\
- 2\rho\sqrt{1-\rho^2} \\
\frac{2\rho^2-1}{\sqrt{2}}
\end{pmatrix} , \\
\ket{\alpha_3} & = \frac{1}{\sqrt{2}} (\ket{R} - \ket{L}) = \frac{1}{\sqrt{2}} \begin{pmatrix}
1 \\
0 \\
-1
\end{pmatrix} .
\end{align*}
The discussion above indicates that the recurrence probability operator should be diagonal when expressed in the basis formed by $\left\{\ket{\alpha_1},\ket{\alpha_2},\ket{\alpha_3}\right\}$. From now on all matrices will be expressed in this basis. For this we will utilize the transition matrices $T$ and $T^\dagger$, where 
$$
T = \begin{pmatrix}
\rho\sqrt{2(1-\rho^2)} & 2\rho^2-1 & \rho\sqrt{2(1-\rho^2)} \\
\frac{2\rho^2-1}{\sqrt{2}} & -2\rho\sqrt{2(1-\rho^2)} & \frac{2\rho^2-1}{\sqrt{2}} \\
\frac{1}{\sqrt{2}} & 0 & -\frac{1}{\sqrt{2}}
\end{pmatrix} .
$$
We denote the matrix $M$ in the $\alpha$ basis as $^\alpha M$.

\begin{figure}
    \centering
    \includegraphics[width=0.6\textwidth]{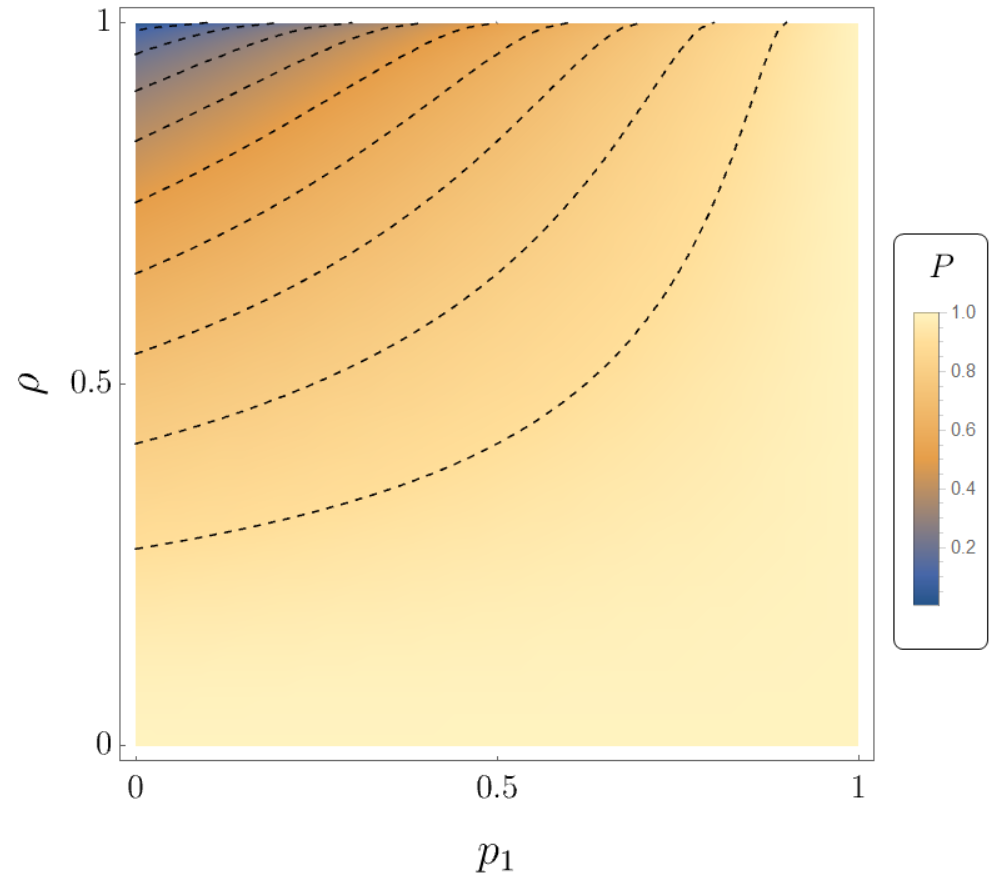}
    \caption{Density plot of the Polya number (\ref{polya:3state}) of the three-state quantum walk as a function of the probability $p_1$ and the coin parameter $\rho$. Dashed lines show the contours of $P = k/10$ for $k=1,\ldots 9$. }
    \label{fig:1}
\end{figure}

\begin{figure}
    \centering
    \includegraphics[width=0.6\textwidth]{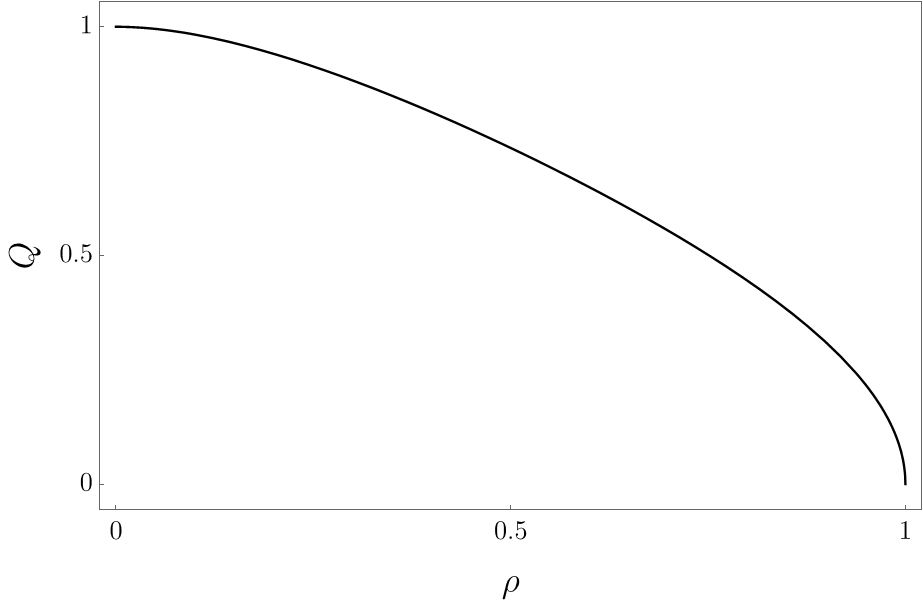}
    \caption{$Q$ (\ref{Q}) as a function of the coin parameter $\rho$. $Q$ represents the minimal value of the recurrence probability for the walk with a given parameter $\rho$.}
    \label{fig:2}
\end{figure}

Turning to the derivation of the Polya number, we begin by expressing the Stieltjes operator through the Fourier transformation (\ref{muc:int}), where $^\alpha U(p)$ is given by
\begin{align*}
^\alpha U(p) & = T\cdot \begin{pmatrix}
e^{i p} & 0 & 0 \\
0 & 1 & 0 \\
0 & 0 & e^{-i p}
\end{pmatrix} \cdot C\cdot T^\dagger \\ 
& = \begin{pmatrix}
2\rho^2-1 & -2\rho\sqrt{1-\rho^2}\cos{p} & -2i \rho\sqrt{1-\rho^2}\sin{p}  \\
-2\rho\sqrt{1-\rho^2} & (1-2\rho^2)\cos{p} & i(1-2\rho^2)\sin{p} \\
0 &  -i \sin{p} & - \cos{p}
\end{pmatrix}.
\end{align*}
The resolvent in the Fourier space is equal to
\begin{align*}
 ^\alpha(I_c - z U(p))^{-1} & = \  \frac{1}{(z-1)(1+z(2-2\rho^2+z) + 2\rho^2 z\cos{p})} \times \\ &  \begin{pmatrix}
 u(z)-2 \rho^2 z \cos p & 2 \rho\sqrt{1-\rho^2} z (\cos p+z) & 2 i \rho \sqrt{1-\rho^2} z \sin p  \\
  2 \rho\sqrt{1-\rho^2} z (z \cos p+1) & w(z) (z \cos p+1)  & -i z v(z) \sin p \\
 -2 i \rho\sqrt{1-\rho^2} z^2 \sin p & i (u(z)+z+1) \sin p & z v(z) \cos p+w(z)  
\end{pmatrix} ,
\end{align*}
where to shorten the formula we have denoted
$$
 u(z)  = z^2(1-2\rho^2)-1,\quad v(z) = 1-2\rho^2+z, \quad w(z) = z(2\rho^2-1)-1. 
$$
The integral in (\ref{muc:int}) can be again evaluated with residues. We will utilize the following integrals (with $n=1,0,-1$)
\begin{align*}
  {\cal I}(n) = \int\limits_0^{2\pi} \frac{dp}{2\pi} \frac{e^{inp}}{(z-1)(1+z(2-2\rho^2+z) + 2\rho^2 z\cos{p})} & = \frac{1}{2\pi i} \oint \frac{x^n dx}{b(x)}, 
\end{align*}
where we have denoted
$$
b(x) = (z-1)(\rho^2 z + (1 + 2(1-\rho^2)z + z^2)x + \rho^2 z x^2). 
$$
The roots of the equation $b(x) = 0$ are
$$
x_\pm = \frac{2 (\rho^2-1)z - z^2 - 1 \pm\sqrt{\left(1 + 2(1-\rho^2)z + z^2\right)^2-4 \rho^4 z^2}}{2 \rho^2 z},
$$
with $|x_-|>1$ for $|z|\leq 1$. Hence, for $n=0,1$ there is only the residue at $x_+$ and we find
\begin{align*}
    {\cal I}(0) & = {\rm Res}\left(\frac{1}{b(x)},x_+\right) = \frac{1}{(z^2-1)g(z)}, \\
     {\cal I}(1) & = {\rm Res}\left(\frac{x}{b(x)},x_+\right) = \frac{ (z+1)g(z) - 1 - 2(1-\rho^2)z - z^2}{2\rho^2 z(z^2-1)g(z)},
\end{align*}
where we have used the notation
$$
g(z) = \sqrt{1 + 2(1-2\rho^2)z + z^2} . 
$$
For $n=-1$ there is an additional residue at $x=0$ and we obtain
\begin{align*}
{\cal I}(-1) & = {\rm Res}\left(\frac{1}{x b(x)},x_+\right) + {\rm Res}\left(\frac{1}{x b(x)},0\right)\\
& = \frac{2\rho^2 z}{(z^2-1)g(z)[(z+1)g(z) - 1 - 2(1-\rho^2)z - z^2]} + \frac{1}{\rho^2z(z-1)}.
\end{align*}
After some algebra, we express the Stieltjes operator in the form
\begin{align}
\label{stl:op}   ^\alpha\mu_c(z) & = \frac{1}{(z-1)g(z)} \begin{pmatrix}
2(\rho^2-1)z - g(z) & \frac{\sqrt{1-\rho^2}(g(z)+w(z))}{\rho} & 0   \\
\frac{\sqrt{1-\rho^2}z(g(z)-v(z))}{\rho} & \frac{w(z)(g(z)-v(z))}{2\rho^2} & 0  \\
0  & 0 & \frac{v(z) g(z)-1-z(z+2-4\rho^2)}{2\rho^2}
\end{pmatrix}.
\end{align}
From (\ref{ac}) we find the first return generating operator
$$
^\alpha a_c(z) = \begin{pmatrix}
(2\rho^2-1)z & -2\rho\sqrt{1-\rho^2} - \frac{4 \rho^3\sqrt{1-\rho^2}(z-1)}{g(z)-v(z)} & 0 \\
-2\rho\sqrt{1-\rho^2}z & 1-2\rho^2 + \frac{2\rho^2(2\rho^2-1)(z-1)}{g(z)-v(z)} & 0  \\
0 & 0 & -\frac{j(z)}{2(\rho^2-1)}
\end{pmatrix}.
$$
Here we have denoted
$$
j(z) = 1 - 2\rho^2 z + z^2 +(z-1)g(z) .
$$
For the product $^\alpha a^\dagger_c(z)\ ^\alpha a_c(z)$ we obtain a diagonal matrix as expected
\begin{equation}
\label{ac2:grover}
    ^\alpha r(z) =  ^\alpha a^\dagger_c(z) \ ^\alpha a_c(z) = \begin{pmatrix}
|z|^2 & 0 & 0 \\
0 & \frac{|j(z)|^2}{4(\rho^2-1)^2} & 0 \\
0 & 0 & \frac{|j(z)|^2}{4(\rho^2-1)^2}
\end{pmatrix} .
\end{equation}
To determine the matrix of the recurrence probability operator in the $\alpha$ basis $^\alpha R$ we have to take $z = e^{it}$ and perform the integral (\ref{rec:op}). The first diagonal entry in (\ref{ac2:grover}) is 1, and so is the result of the integral, corresponding to the fact that for the $\ket{\alpha_1}$ state the walker returns to the original site with certainty. Then we are left with evaluating 
$$
Q = \int\limits_0^{2\pi} \frac{dt}{8\pi(\rho^2-1)^2} |j(e^{it})|^2. 
$$
We have to carefully express $g(e^{it})$ and $j(e^{it})$. We find the following
$$
g(e^{it}) = \left\{\begin{array}{cc}
    e^{\frac{i t}{2}} \sgn{(\sin{t})} \sqrt{2(\cos{t} + 1 -2\rho^2)} , & \cos{t}\geq 2\rho^2 - 1 \\ \\
     -i e^{\frac{i t}{2}} \sqrt{-2(\cos{t} + 1 -2\rho^2)} , & \cos{t}< 2\rho^2 - 1
\end{array} \right. .
$$
Hence, for $t\in (0,\arccos(2\rho^2-1))\cup (2\pi-\arccos(2\rho^2-1),2\pi)$ , we obtain
$$
j(e^{it}) = 2 e^{it}\left(\cos{t} - \rho^2 + i \sgn{(\sin{t})}\sin(t/2) \sqrt{2(\cos{t}+1 -2\rho^2)} \right),
$$
which has a constant modulus square
\begin{align*}
|j(e^{it})|^2 = 4(\rho^2-1)^2 .
\end{align*}
The contribution to $Q$ is then proportional to the length of the interval and is found to be
$$
Q_1 =  \frac{\arccos(2\rho^2-1)}{\pi}.
$$
Turning to the case $\cos{t}< -2\rho^2-1$, i.e. the interval $t\in(\arccos(2\rho^2-1),2\pi-\arccos(2\rho^2-1))$, we find
$$
j(e^{it}) = 2 e^{it}\left(\cos{t} - \rho^2 +\sin(t/2) \sqrt{-2(\cos{t}+1-2\rho^2)} \right).
$$
The resulting integral can be evaluated directly
\begin{align*}
Q_2 = &  \int\limits_{\arccos{(2\rho^2-1)}}^{2\pi-\arccos(2\rho^2-1)} \frac{dt}{8\pi(\rho^2-1)} |j(e^{it})|^2 \\
 = &  \int\limits_{\arccos{(2\rho^2-1)}}^{2\pi-\arccos(2\rho^2-1)} \frac{dt}{2\pi(\rho^2-1)} \left(\cos{t} - 1 +\sin(t/2) \sqrt{-2(\cos{t}+1-2\rho^2)} \right)^2 \\
 & =  \rho\frac{2(1+2\rho^2)\sqrt{1-\rho^2}- \rho(2+\rho^2)\arccos(2\rho^2-1)}{\pi} .   
\end{align*}
In summary, we find
\begin{equation}
\label{Q}
 Q = Q_1 + Q_2 = \frac{2 \rho \left(2 \rho^2+1\right) \sqrt{1-\rho^2}+\left(1-4 \rho^2\right) \arccos\left(2 \rho^2-1\right)}{\pi  \left(\rho^2-1\right)^2} .   
\end{equation}
The matrix of the recurrence probability operator in the $\alpha$ basis is then given by
$$
^\alpha R = \begin{pmatrix}
1 & 0 & 0 \\
0 & Q & 0 \\
0 & 0 & Q
\end{pmatrix} .
$$
This means that the Polya number for the three-state walk with the coin (\ref{coin}) depends on the probability $p_1$ to be initially in the $\ket{\alpha_1}$ state
$$
p_1 = |\braket{\alpha_1}{\psi_c}|^2,
$$
and the coin parameter $\rho$. The resulting formula is
\begin{equation}
\label{polya:3state}
    P(\psi_c) = p_1 + Q(1-p_1)  .
\end{equation}
The site recurrence probability thus ranges between 1 and $Q$ (\ref{Q}). In particular, for the Grover walk corresponding to $\rho = \frac{1}{\sqrt{3}}$, we find
$$
P(\psi_c) = p_1 + \frac{10\sqrt{2} - 3\arccos{(-1/3)}}{4\pi} (1-p_1) \stackrel{.}{=} 0.67 + 0.33 p_1 .
$$
We illustrate our results in Figures~\ref{fig:1} and \ref{fig:2}. Figure~\ref{fig:1} shows the site recurrence probability (\ref{polya:3state}) in dependence of $\rho$ and $p_1$. In Figure~\ref{fig:2} we display the value of $Q$ (\ref{Q}) as a function of the coin parameter $\rho$. 

\section{State recurrence of a three-state walk on a line}
\label{sec4}

Let us now briefly comment on the state recurrence \cite{grunbaum_recurrence_2013}. In this case we are interested in the return to the exact initial state, so the orthogonal projector (\ref{pi0}) has the form
$$
\Pi_0 = \ketbra{0}{0}\otimes \ketbra{\psi_c}{\psi_c}.
$$
The recipe to determine the state recurrence probability $S(\psi_c)$ is similar to the site recurrence, but instead of the operator valued generating functions we will deal with scalars. The matrix element of the Stieltjes operator (\ref{stl:op}) with the initial coin state $\ket{\psi_c}$ yields the generating function for return amplitudes without prior monitoring
$$
\mu_{\psi_c}(z) = \braketA{\psi_c}{\mu_c(z)}{\psi_c}.
$$
The generating function for the first arrival amplitudes is then given by \cite{grunbaum_recurrence_2013}
$$
a_{\psi_c}(z) = 1 - \frac{1}{\mu_{\psi_c}(z)}.
$$
Finally, the state recurrence probability is obtained by a formula analogous to (\ref{rec:op})
\begin{equation}
S(\psi_c) = \int\limits_0^{2\pi} \lvert a_{\psi_c}(e^{i t})\rvert^2 \frac{dt}{2\pi}.
\label{state:rec}
\end{equation}
In contrast to the site recurrence, which was tractable, the integrand in (\ref{state:rec}) is usually a rather complicated function and we rely on numerical integration. Below we present plots with numerical results for several initial states.

\begin{figure}
    \centering
    \includegraphics[width=0.6\textwidth]{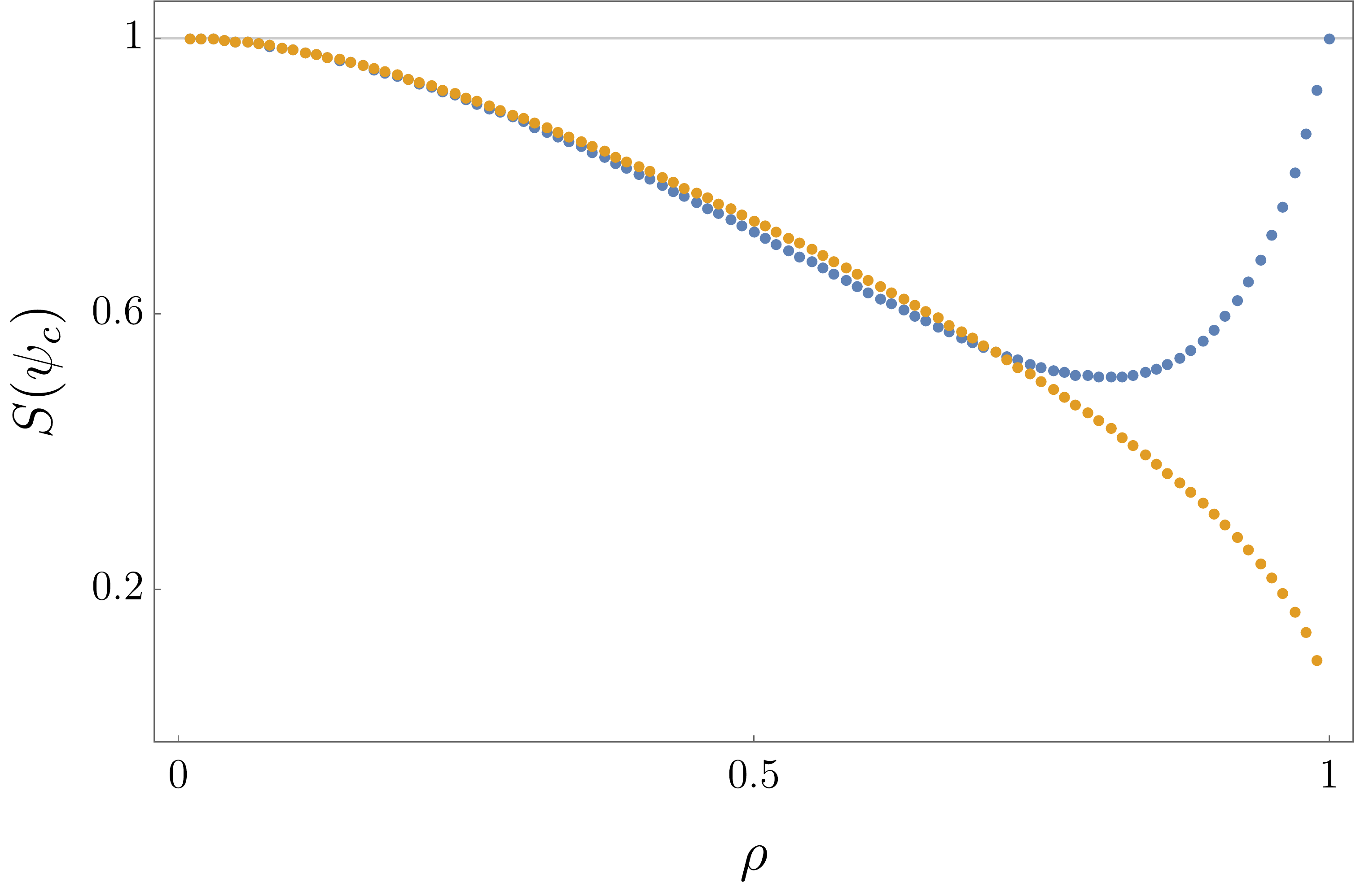}
    \caption{State recurrence probability for the state $\ket{\alpha_1}$ (blue dots) and $\ket{\alpha_2}$ and $\ket{\alpha_3}$ (orange dots)  in dependence on $\rho$.}
    \label{fig:3}
\end{figure}

For Figure~\ref{fig:3} we consider the basis states $\ket{\alpha_i}$ and plot the state recurrence probability as a function of the coin parameter $\rho$. For $\ket{\alpha_2}$ and $\ket{\alpha_3}$ the integral in (\ref{state:rec}) can be evaluated analytically and we find that the state recurrence is equal to the site recurrence. In fact, numerical simulations reveal that for the initial state $\ket{\alpha_3}$ the walker always returns to the origin in the $\ket{\alpha_3}$ state. However, for $\ket{\alpha_2}$ the situation is more complicated - the walker returns to the original site in a superposition of $\ket{\alpha_1}$ and $\ket{\alpha_3}$ with different weights in every time step. Nevertheless, the state recurrence probability equals (\ref{Q}). For $\ket{\alpha_1}$ the state recurrence probability is smaller than site recurrence probability (which equals 1), except for the boundary cases of $\rho =0, 1$.

\begin{figure}
    \centering
    \includegraphics[width=0.6\textwidth]{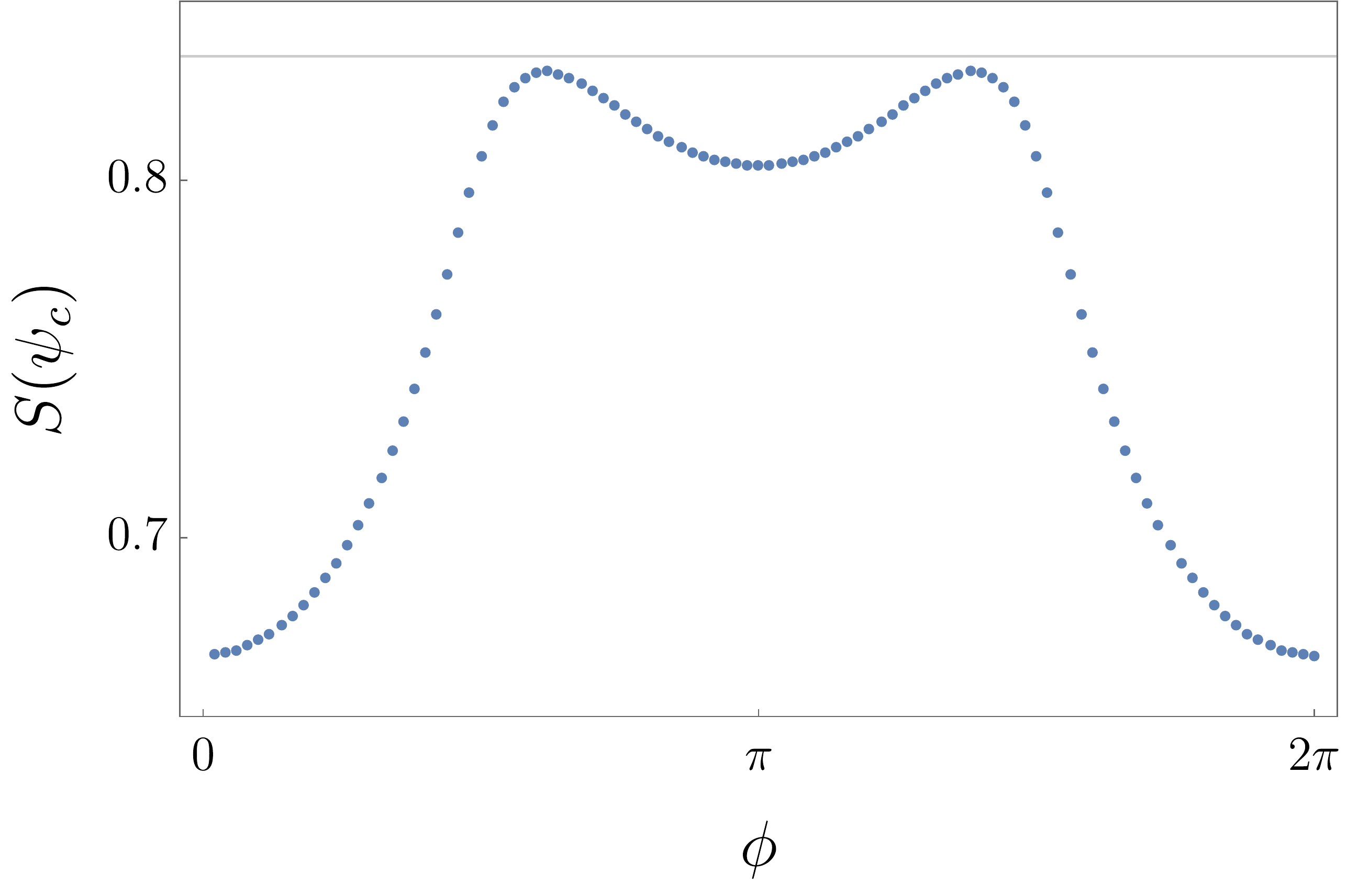}
    \caption{State recurrence probability of the Grover walk ($\rho =1/\sqrt{3}$) for the state $\frac{1}{\sqrt{2}} (\ket{\alpha_1} + e^{i\phi}\ket{\alpha_2})$ as a function of the angle $\phi$. }
    \label{fig:5}
\end{figure}

\begin{figure}
    \centering
    \includegraphics[width=0.6\textwidth]{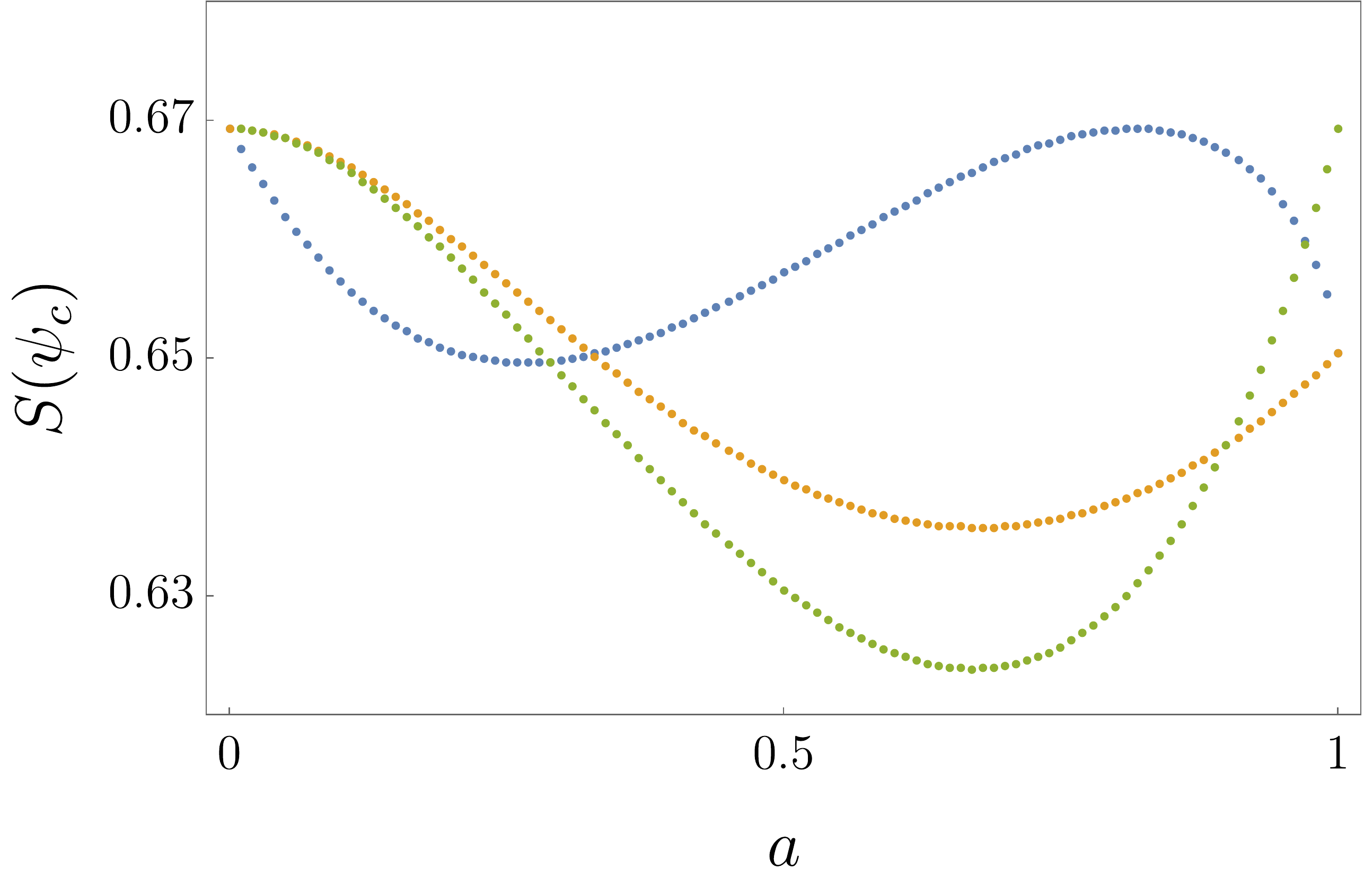}
    \caption{State recurrence probabilities of the Grover walk ($\rho =1/\sqrt{3}$) for the states $ a\ket{\alpha_1} + \sqrt{1-a^2}\ket{\alpha_2}$ (blue dots), $ a\ket{\alpha_1} + \sqrt{1-a^2}\ket{\alpha_3}$ (orange dots) and $ a\ket{\alpha_2} + \sqrt{1-a^2}\ket{\alpha_3}$ (green dots) as a function of $a$.}
    \label{fig:6}
\end{figure}

Figures~\ref{fig:5} and \ref{fig:6} focus on the three-state Grover walk ($\rho = \frac{1}{\sqrt{3}}$). In Figure~\ref{fig:5} we consider the initial state $\frac{1}{\sqrt{2}} (\ket{\alpha_1} + e^{i\phi}\ket{\alpha_2})$ and plot $S$ as a function of the angle $\phi$. Note that for superpositions $\frac{1}{\sqrt{2}} (\ket{\alpha_1} + e^{i\phi}\ket{\alpha_3})$ and $\frac{1}{\sqrt{2}} (\ket{\alpha_2} + e^{i\phi}\ket{\alpha_3})$ the angle $\phi$ does not affect the state recurrence probability due to the block-diagonal form of the Stieltjes operator (\ref{stl:op}). Figure~\ref{fig:6} shows state recurrence probability for superposition states $a\ket{\alpha_1} + \sqrt{1-a^2}\ket{\alpha_2}$, $a\ket{\alpha_1} + \sqrt{1-a^2}\ket{\alpha_3}$  and $a\ket{\alpha_2} + \sqrt{1-a^2}\ket{\alpha_3}$ as a function of $a$. 

\begin{figure}
    \centering
    \includegraphics[width=0.6\textwidth]{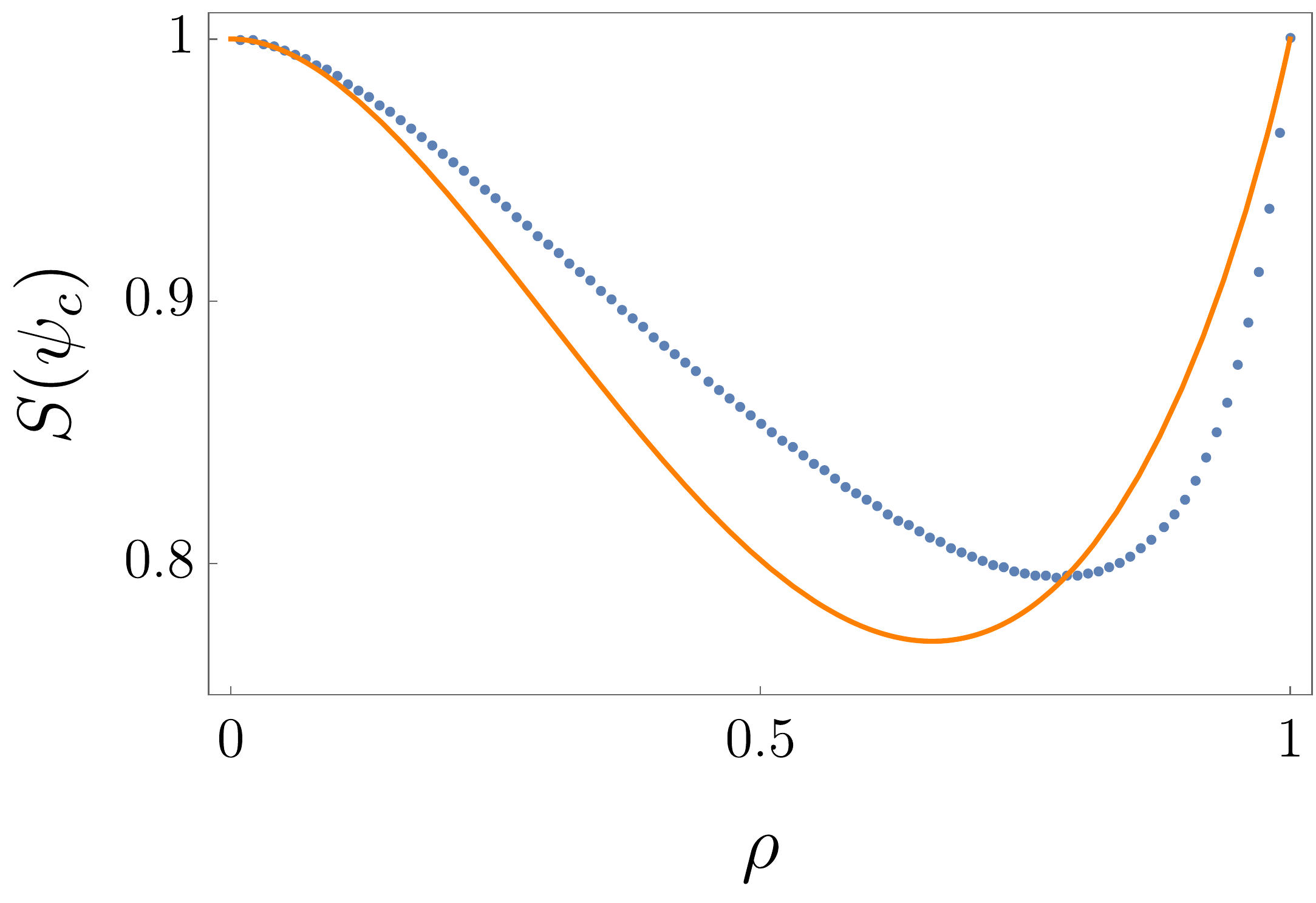}
    \caption{State recurrence (blue dots) versus site recurrence (orange curve) for the initial coin state (\ref{init:spec}).}
    \label{fig:7}
\end{figure}

Finally, we point out that the three-state walk shows similar paradoxical behavior as reported in \cite{gruenbaum2014} for a walk on a half-line or some 2D quantum walks, where the state recurrence probability can be greater than the site recurrence probability. In Figure~\ref{fig:7} we consider the initial coin state 
\begin{equation}
    \label{init:spec}
    \ket{\psi_c} = \sqrt{\frac{1-\rho^2}{2}}\ket{R} + \rho\ket{S} + \sqrt{\frac{1-\rho^2}{2}}\ket{L} = \rho\ket{\alpha_1} -\sqrt{1-\rho^2}\ket{\alpha_2},
\end{equation}
which is one of the eigenvectors used in the construction of coin operator \cite{stefanak2012}. The plot indicates that the site recurrence probability, depicted by the orange curve, is smaller than the state recurrence probability up to $\rho\approx 0.79$.

\section{Conclusions}
\label{sec5}

The presented results demonstrate that in contrast to the simple two-state quantum walks the site recurrence of the three-state model depends on the initial state. For the selected one-parameter family of coins we were able to derive the site recurrence probability in closed form. Extension to an arbitrary 3x3 unitary coin seems to be intractable due to the complexity and large number of free parameters. Nevertheless, for all three-state walks there will be an initial coin state with Polya number equal to one, namely the state given by $C^{-1}\ket{S}$. Indeed, such state will remain on the initial position after the first step, and thus is absorbed with certainty. It is an open question if such behaviour of site recurrence is possible for models without the staying put option in the shift operator, e.g. in quantum walks driven by Wigner rotation matrices of order $2j+1$ studied in \cite{miyazaki2007,Bezdekova2015} for the case of half-integer $j$.

As discussed in Section \ref{sec4}, obtaining closed formulas for state recurrence probability was possible only for particular initial coin states due to the complexity of the involved integrals. The same applies to both site and state recurrence for quantum walks on higher dimensional lattices, since in such cases already the evaluation of the Stieltjes operator through multidimensional Fourier transform cannot be easily reduced to calculation of residues as in the one-dimensional case. Nevertheless, the formulas allow to obtain approximations through numerical evaluation. It would be interesting to find if there exists site recurrent initial conditions e.g. for 2D quantum walks with Grover coin and its extensions. Additional twist can come from considering coined quantum walks combined with some form of classical randomness resulting in iterated open quantum evolution, e.g. quantum walks on dynamically percolated lattices \cite{Kollar2012,Kollar2014,Kollar2014EPJP}. Recent extension of monitored recurrence to arbitrary iterated closed operators on Banach spaces \cite{grunbaum2018} and quantum Markov chains \cite{grunbaum_quantum_2020} allows to investigate such scenarios. 
 
\ack

M\v S is grateful for financial support from RVO14000 and ”Centre for Advanced Applied Sciences”, Registry No. CZ.02.1.01/0.0/0.0/16 019/0000778, supported by the Operational Programme Research, Development and Education, co-financed by the European Structural and Investment Funds. 

On a personal note, I want to express my gratitude to prof. Igor Jex for years of stimulating discussions and mediation of a myriad of personal contacts with excellent researchers worldwide. Live long and prosper!

\section*{References}
\bibliography{recbib}
\bibliographystyle{unsrt}

\end{document}